# Genesis of thunderstorm ground enhancements

*A.Chilingarian, G.Hosepyan, T.KArapetyan, and B.Sargsyan*

*Yerevan physics institute, Alikhanyan Brothers 2, Yerevan, Armenia, 0036*

**Abstract**

Proceeding from a stormy day of 22 September 2022, when 7 thunderstorm ground enhancements occurred (TGEs, 3 of them very large), we perform an analysis of the most important conditions, on which depend the origination of the large particle fluxes in the thunderous atmosphere. Among these conditions are the near-surface electric field (NSEF), graupel fall, and lightning activity. We estimate the intensity of the largest particle flux of 1,25 mln gamma rays hitting each square meter of surface on Aragats with energy spectra extended up to 70 MeV.
Only one TGE from 7 meets the conditions to recover the electron energy spectrum; the fraction of electrons with energies above 10 MeV relative to the gamma ray flux reaches 45%.
By carefully examining the graupel fall, we demonstrate that the lower dipole, which accelerates electrons, is formed by the main negative and lower positively charged regions. The lower dipole decays with a graupel fall that coincides with TGE terminations (usually by a lightning flash).

### 1. Introduction: Aragats Solar Neutron Telescope (ASNT) quality check

Summer 2022 on Aragats (usually July-August on Aragats research station, 3200 m a.s.l.) was dry and hot. In 3 months, there were only 6 thunderstorms, as we can see in Fig.1 by disturbances of the near-surface electric field (NSEF, black time-series) and by lightning flashes, (green lines at the top of Figure). The outside temperature was higher than usual, exceeding 15C half of the time. On the time series of count rate of particle detector (blue curve), impulsive enhancements were very small, suddenly giving enormous large peaks on 22 September, when 3 huge particle flux enhancements were registered.

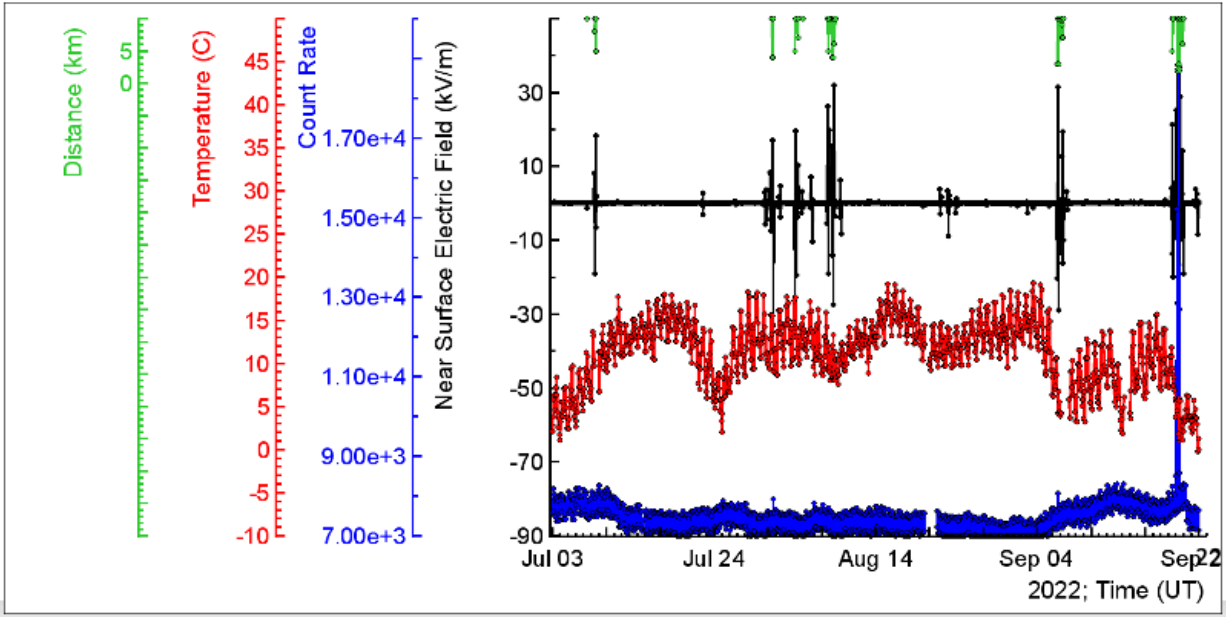

Figure 1. Summer 2022: disturbances of the NSEF, black; time series of 1-minute count rates of STAND3 plastic scintillator of 1 m$^2$ area and 3 cm thickness, blue; outside temperature, red; distances to lightning flashes, green.

Thunderstorm ground enhancements (TGEs) occurred during severe thunderstorms when the electric field in the atmosphere exceeds critical energy for unleashing a runaway process [1], during which the free electrons from the numerous extensive air showers (EASs) developing in the atmosphere, multiplied, accelerated and form electron-gamma ray avalanches reaching the earth's surface [2,3]. In Fig.2 we present 4 episodes of TGE registration the by Aragats research station [4,5].

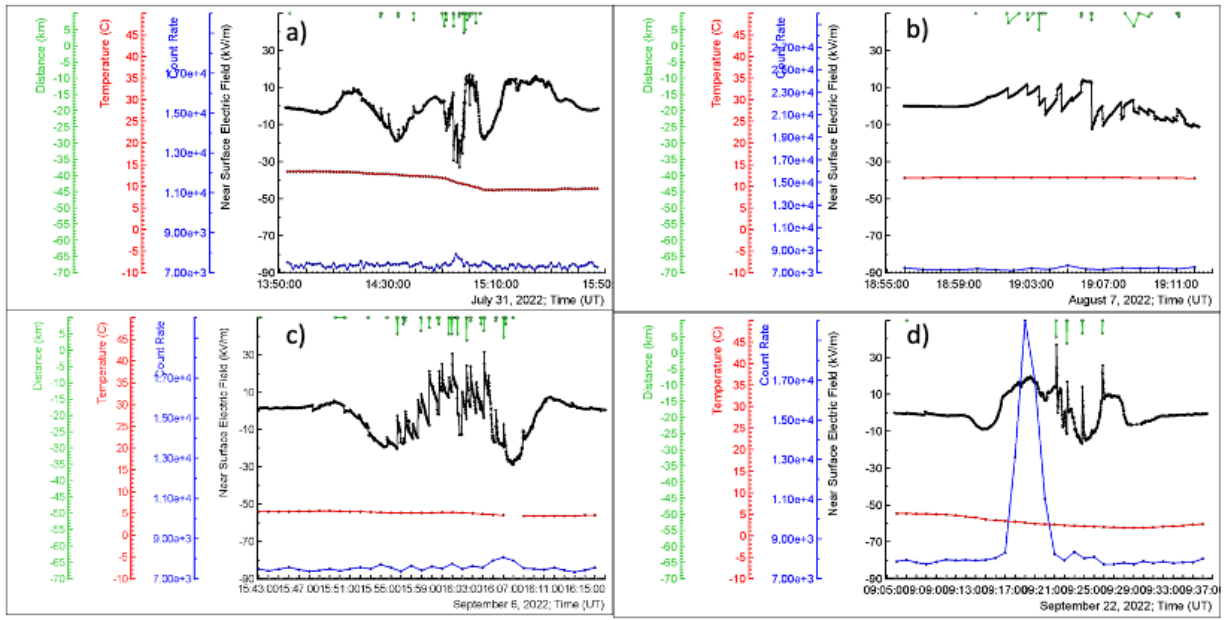

Figure 2. TGE events occurred on Aragats from July to September 2022. The description of colored curves is the same as in Fig. 1.

The particle flux enhancement during very few "summer TGEs" shown in Fig 2 a-c never exceeds 8%, and the corresponding peak significance measure in the number of standard deviations above fair-weather value, never exceeds 10 . And suddenly, on September 22, during an ordinary storm, in 5 hours detectors registered 3 record enhancements, the largest of which at 9:20 demonstrates ever largest enhancements of 157% and 148  in 1-minute time series (STNAD3 detector's "1000" coincidence, Fig 2d).

To understand environments leading to such an enormous enhancement we study this event in all detail performing necessary reliability checks.

Among numerous particle detectors operated on Aragats, the most important is Aragats solar neutron telescope (ASNT), the only spectrometer in the TGE research measuring electron and gamma ray energy spectra separately. First of all, we check the uniformity of operation of 8 scintillators of TGE, calculating the background count rates before and after the event. In Fig. 3 we show rather good uniformity of the count rates, varying within 10% for the thick scintillators, and within 5% for the thin ones. During TGE the difference is larger reaching 30% for second and third thick scintillators (blue and red) due to slightly different energy thresholds. However, we present energy spectra only for energies above 10 MeV, thus, the difference between count rates of four scintillators is not too essential.

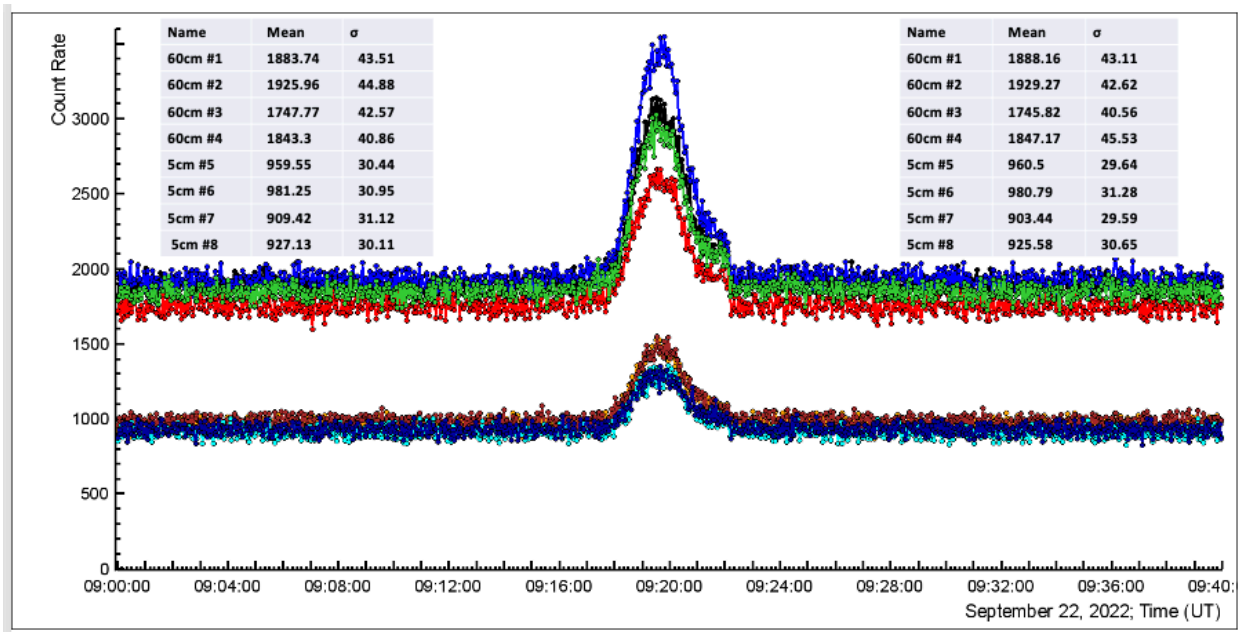

**Figure 3. 2s time series of count rates of 8 ASNT scintillators (four 5 cm thick "veto" scintillators, and four 60 cm thick spectrometric scintillators). In the insets mean values and variances of the count rates before and after TGE are depicted.**

## 2. The largest TGE event on 22 September 2022

The big advantage of the ASNT spectrometer is that all possible coincidences in the ASNT scintillators are measured and stored separately. In Fig. 4 we present the time series of count rates of the sum of 4 thick scintillators with and without the veto option.

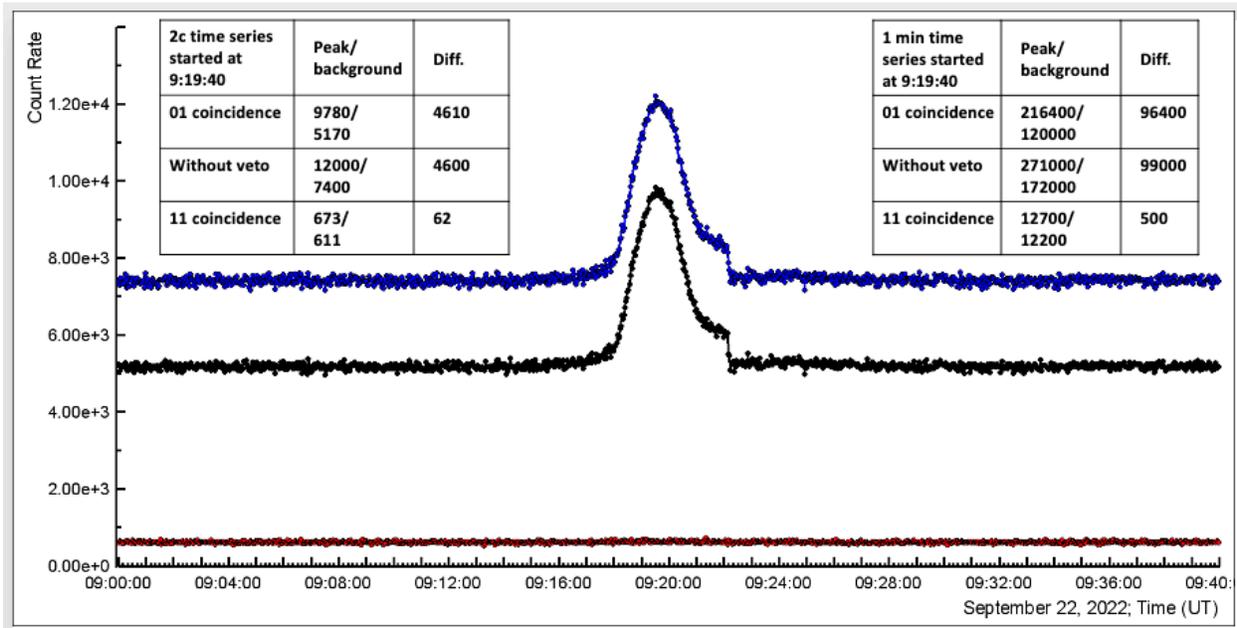

**Figure 4.** 2s time series of count rates of ASNT coincidences: "01", mostly gamma rays, black; "11", mostly electrons, red; and count rate of all particles registered in the lower 60 cm thick scintillator, blue. In the insets, the peak and background values of different ASNT coincidences are shown and the difference (TGE flux) is calculated for the 2s (left inset) and 1 minute (right inset) time series of the count rates.

As we can see in Fig. 2d, TGE started at 19:17 when NSEF was near zero and develops along with the enhancement of NSEF. At the maximum of TGE flux at 9:19:30, the NSEF reaches 20 kV/m, then TGE declines with attenuation of the NSEF until normal polarity lightning flash abruptly terminates it at 9:22:10.

In Fig.4a we see that at 9:19:40 electron detection was negligible, and we can claim that the electron flux was too small to be recovered (see the left inset). However, a 2s time span is not enough to register highly variable electron flux, thus, we summarize thirty 2s counts to obtain a 1-minute count rate (see the right inset in Fig.4). In 1-minute we have 500 electron candidates (coincidence "11"), in turn, number of gamma ray candidates is 96400 (coincidence "01"). Thus, it will be very difficult, to disentangle the 0.5% fraction during the energy recovery procedure. We need at least a few percent shares of electron candidates to start a rather complicated energy recovery procedure (see details in [6]).

Nonetheless, we can see a notice of electron content in Fig. 5. In Fig 5a we show the 1-minute time series of "11" coincidence, and in Fig. 5b we show the energy release histogram in 5 cm thick upper scintillator of ASNT.

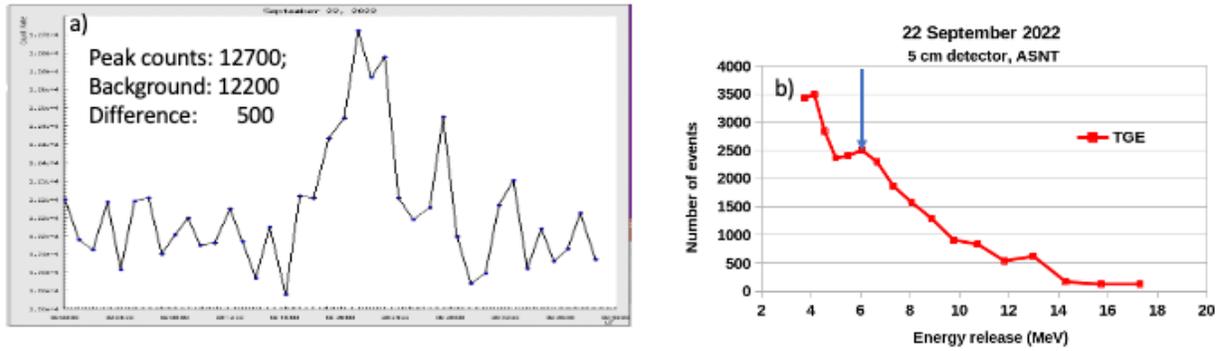

Figure 5. a) 1-minute time series of the "11" coincidence of ASNT; b) the energy release histogram in the 5-cm thick upper scintillator of ASNT.

By the blue arrow, we point to a small peak in the energy release distribution. As electrons leave ≈1.8 MeV energy release due to ionization in every cm of the scintillator, we expect a peak from electrons in the 6-9 MeV interval. The distribution of the energy losses by gamma rays follows exponential law and does not exhibit any peaks. Thus, this small deviation from the exponent is due to a small portion of electrons in the TGE flux, not enough for the recovery of the electron energy spectrum. Therefore, we recover only gamma ray energy spectrum with another spectrometer network operated on Mt. Aragats, namely, a network of large (12 x 12 x28 cm) NaI (TL) spectrometers with the energy threshold of 300 KeV (see details in [7]). As we can see in Fig. 6 energy spectra exhibit a "knee" feature) around 9-10 MeV. Correspondingly the fitted spectra depend on 5 parameters, analogically as to the primary cosmic ray spectrum measured by MAKET ANI surface array [8]:

$J_{TGE} = A*E^{-\gamma} (1 + (E/E_{knee})^E )^{\Delta\gamma/\epsilon}$,

where $\gamma$ varies from 0.98 to 1.06, $\Delta\gamma$ – from 2.46 to 3.17, and the sharpness of the knee $\epsilon$ from 2.34 to 2.66.

Parameters of intense energy spectra measured on 22 September are summarized in Table 1 Spectra have a rather complicated shape before 10 MeV due to Radon progeny gamma radiation [9], ambiguities in detector response function calculation at low energies, and, possibly some unknown modulation effects in the atmospheric electric field. However, the differential energy spectra in the energy region 10 – 50 MeV follow the exponential law with very high accuracy. The "knee", the spectrum turnover point occurred around 10 MeV for the much intense TGE that occurred 9:18-9:22, and 6-8 MeV for the other 2 TGEs. All spectra are rather smoothly hardening at high energies, changing from (1-1.6) to (3-4). As we mentioned above the TGE, which started at 9:18 and was terminated by a lightning flash at 9:21:10, thus, the first minute 9:18 - 9:19 and the last minute 9:21- 9:22 spectra are much weaker than at 9:18:9:21. Gamma ray energy spectrum measured at minute 9:19-9:21 (Fig. 6b) is immensely intensive and energetic (extended to 70 MeV). In one minute ≈1,25 mln gamma rays with energies above 0.3 MeV hit every square meter of the earth's surface covering several square kilometers on the ground, see Fig.7.

Table 1. Parameters of the 5-parametric fit of the TGE gamma ray energy spectra (parameter errors are shown in the legends in Figs. 6 and 14)

| Time 22/09 2022 | $\gamma_1$ | $\gamma_2$ | A/100,000 | $E_{knee}$ | Sharpness |
|---|---|---|---|---|---|
| 4:22 | 1.61 | 2.97 | 2.63 | 6.15 | 6.98 |
| 4:46 | 1.53 | 3.01 | 1.73 | 7.18 | 5.89 |
| 4:47 | 1.49 | 3.37 | 1.79 | 8.14 | 2.76 |
| 9:18 | 0.98 | 3.44 | 1.45 | 8.37 | 2.66 |
| 9:19 | 1.06 | 3.93 | 3.81 | 10.13 | 2.45 |
| 9:20 | 1.05 | 3.72 | 2.71 | 9.19 | 2.54 |
| 9:21 | 1.01 | 4.18 | 0.95 | 10.00 | 2.34 |

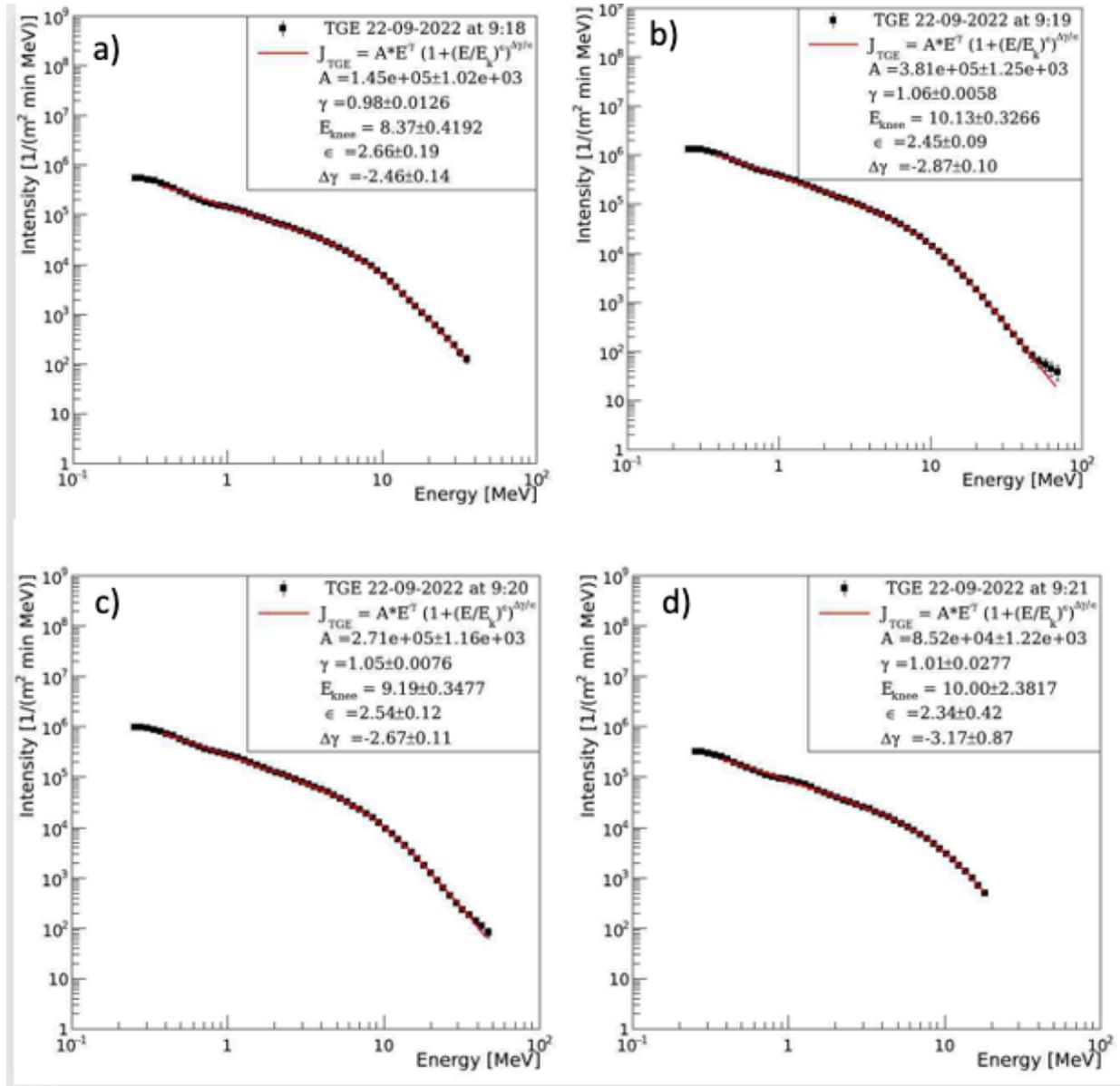

**Figure 6.** Differential energy spectra recovered from the NaI(TLL) spectrometer N2. Parameters of 5-parametric fir are shown in the legend of each frame.

The enhancement of the count rate overhead the fit line at energies above 50 MeV (Fig. 6b and lesser in Fig. 6c) can be explained by the MOS process, which contribution becomes essential at highest energies [10].

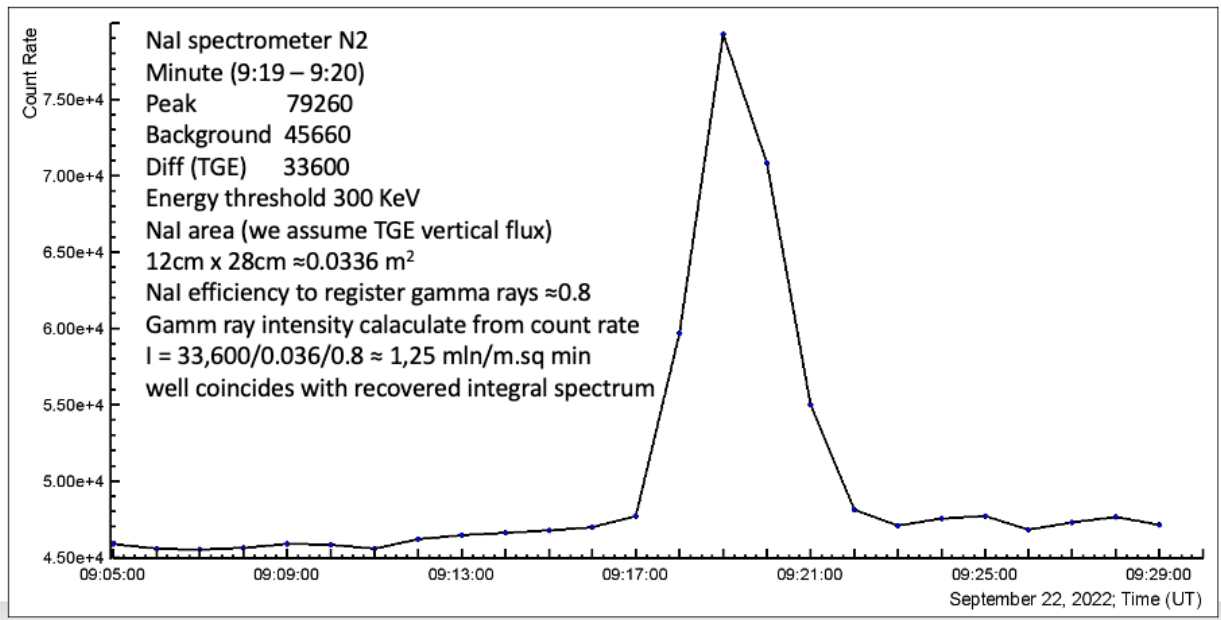

**Figure 7. Time series of 1-minute count rate of NaI spectrometer N2. In the inset, we show the recovered from the count rate the integral intensity of the TGE.**

The electron energy spectrum should extend even to higher energies, possibly until 100 MeV at the exit from the strong accelerating electric field at ≈200 m above the ground. Electrons do not reach the ground in sizable quantities to allow their energy spectrum recovery. As we can see in Fig. 8 the distance to the cloud base during TGE was ≈200 m. If we assume that the electric field is prolonged only in the cloud, it is too large a distance even for > 50 MeV electrons to reach the ground in sizable quantities.

Nonetheless, if electrons are accelerated and multiplied in the lower dipole formed by the main negative region in the middle of the cloud and transient positive charged region (LPCR) we can expect that the electric field is extended lower than the cloud base and some electrons reach the ground. LPCR is sitting on fallen graupel, thus, we can expect a continuous acceleration of electrons in the capacitor with a "moving" second plate until all graupel reaches the earth's surface. The graupel fall can be noticed by panoramic cameras monitoring skies above Aragats, see Fig.9. The characteristic specks on the glass of the camera are graupel, the identification was made by the station staff, see Figs 11 and 12 in [11].

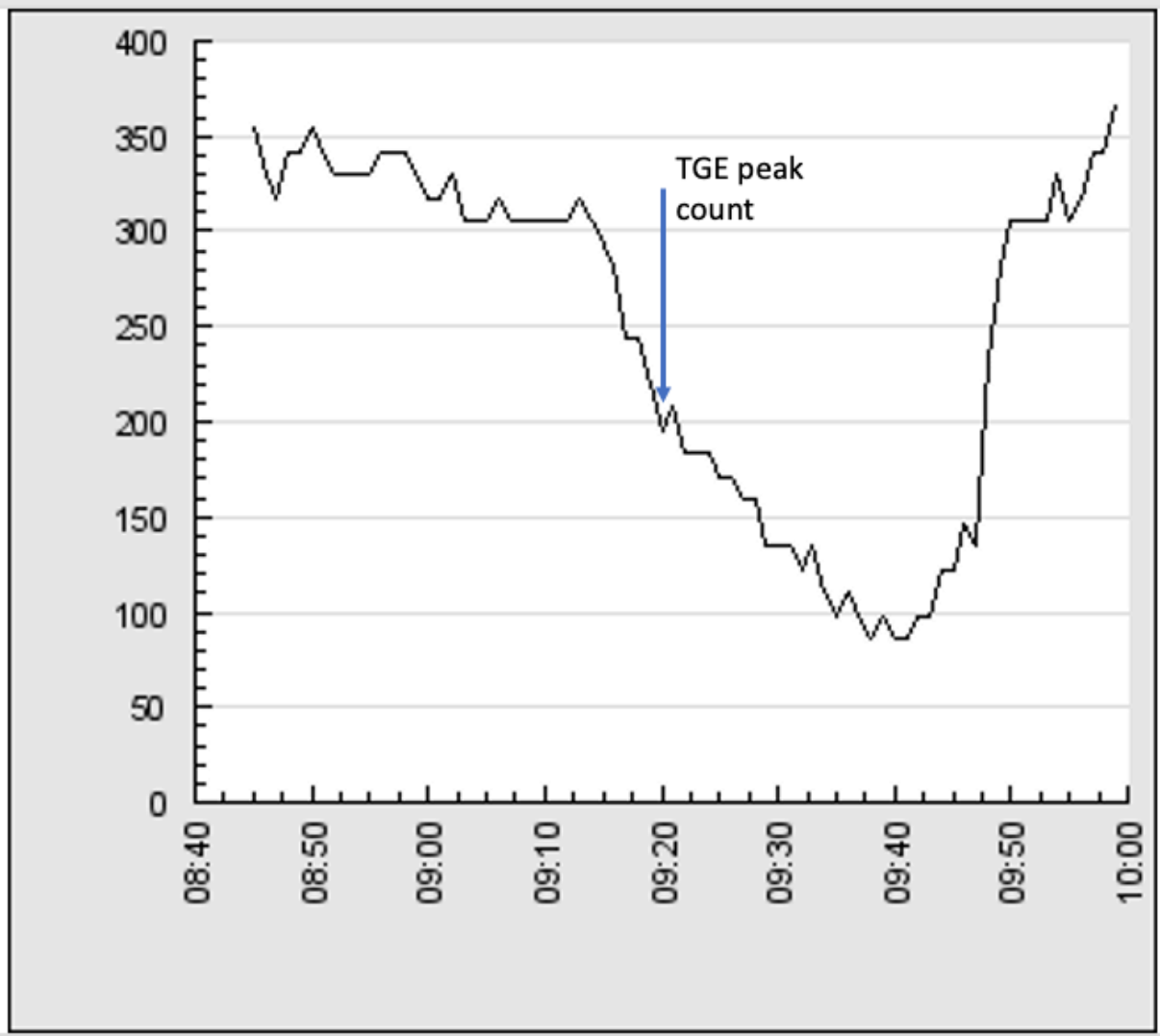

Figure 8. Time series of distances to the cloud base during TGE with peak count at 9:20.

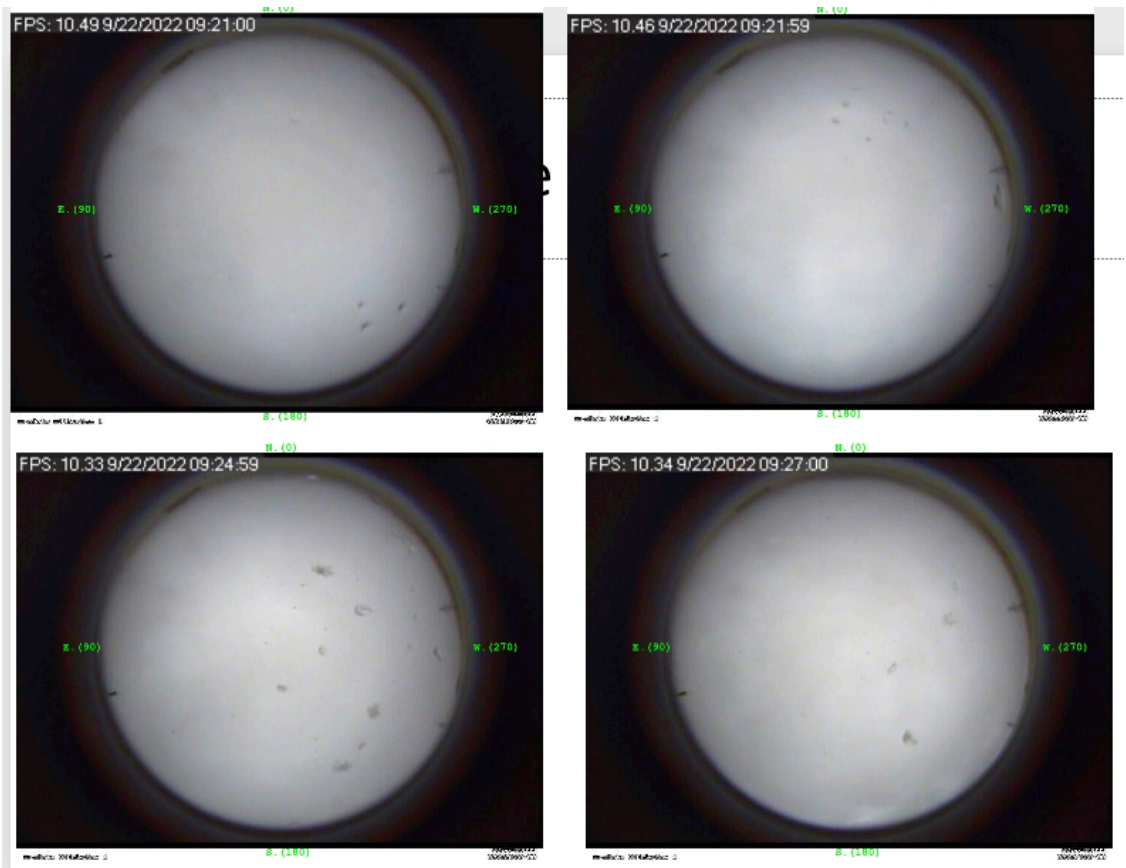

**Figure 9. The panoramic shots of the skies above Aragats just after the TGE. The specks on the photos are identified with graupel fallen on the glass of the camera.**

As we can see from the times of the camera shots the graupel fall starts just after the maximum of the TGE intensity and prolongs for 6 minutes.

3. **The electron energy spectrum recovery from the large TGE that occurred on 22 September 2022 at 4:22.**

On the morning of 22 September during half-of-hour, another 2 large short-duration TGEs occurred (local time 8:22 and 8:47, duration ≈1 and 2.5 minutes), see Fig.10. In spite of that both were very close by the amplitude of the flux enhancement, and by the normal polarity of lightning flashes that terminated them, the polarity of the NSEF was opposite. In Fig. 11 we can see that the first TGE of this remarkable day, which occurred during a deep negative electric field (≈-20 kV/m) was terminated by a lightning flash just after reaching the maximum flux (the flux decreases by 34% in 2s), and the second during positive NSEF (+15 kV/m) at the end of the TGE development (the flux decreases by ≈10% in 2s). These 2 scenarios of the lower dipole emergence between the main negative layer and its mirror in the earth, and between the same main negative and LPCR were described in detail in [12].

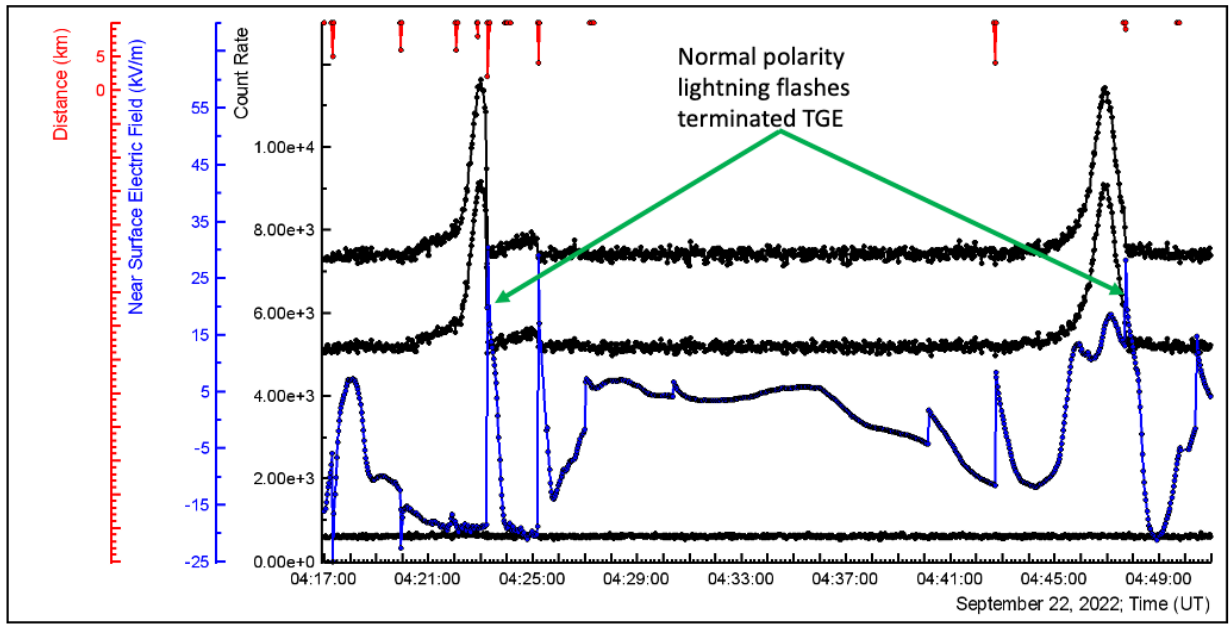

**Figure 10. Black from top to bottom: 2s time series of ASNT detector count rates, without veto, and with a veto to charged particles ("01" coincidence), and mostly electrons ("11" coincidence); blue – disturbances of NSEF with abrupt increases corresponding to the lightning flashes; red distances to lightning flash. Green arrows are denoted normal polarity lightning flashes, which terminate TGEs at 4:23:12 (distance to the flash 3.5 km), and on 4:47:42 (distance to the flash – 8.5 km).**

In Fig. 11a and 11b, we show zoomed versions of TGE events shown in Fig. 11 now with different colors and in percent of enhancement relative to the fair-weather mean count rate. During the first TGE that occurred from 4:22:14 to 4:23:12, we can notice a ≈10% enhancement of the count rate of "11" coincidence, the green curve (above the yellow line). Thus, we can expect a sizable number of electrons. In the second 2-minute long TGE, Fig.11a, there is no enhancement of the "11" coincidence above the yellow line, and consequently, no electron reaches the detector.

The next step to prove that the TGE electrons reach the detector is the examining of the energy release histogram in the upper 5cm thick scintillator, see Fig.12. This time, we simulate the energy release of gamma rays in the 5-cm thick scintillator using the gamma ray energy spectra recovered by "01" coincidence in the 60cm thick "spectrometric" scintillator (we join three 20s energy release histograms to recover 1-minute gamma ray and electron differential energy spectra).

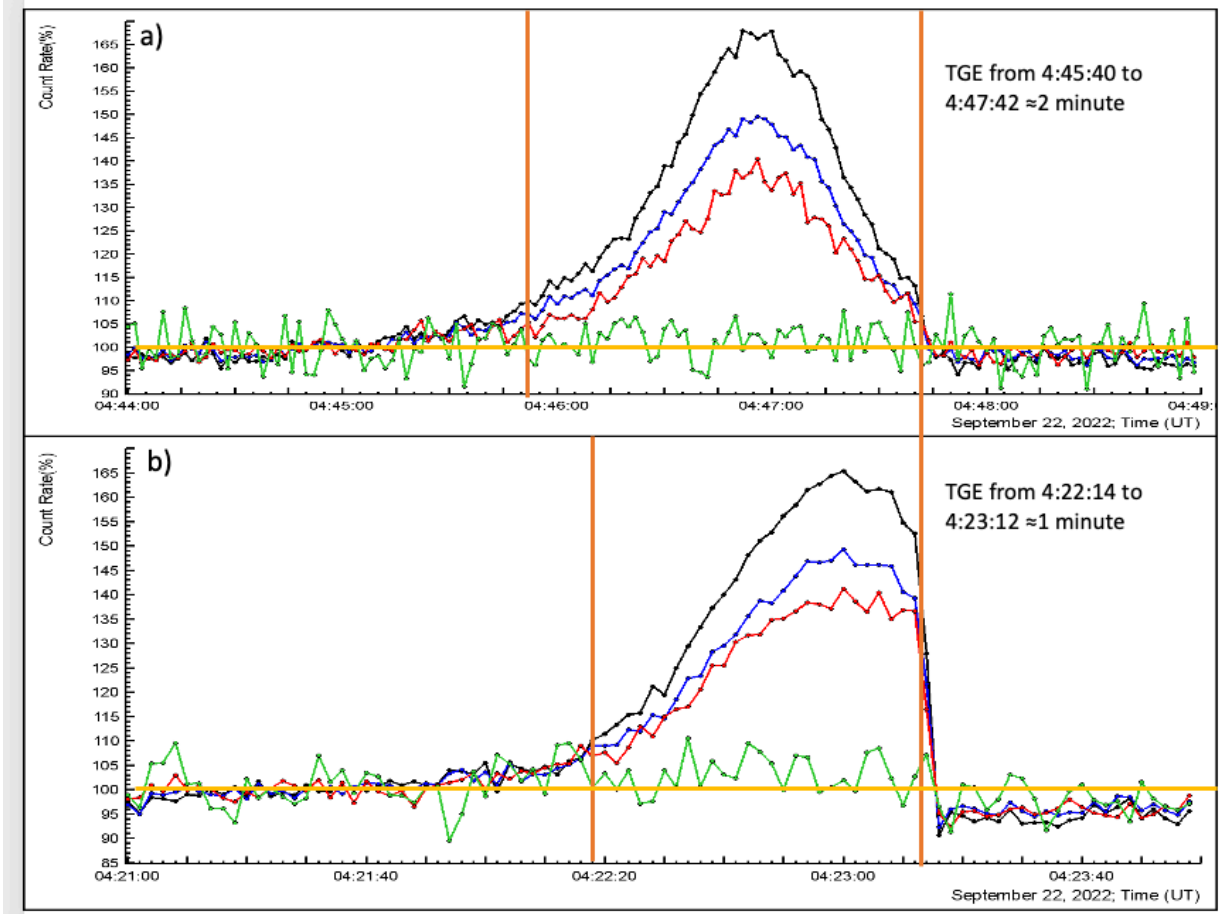

Figure 11. a) 2s time series of ASNT detector count rates in percent to the fair-weather mean with a veto to charged particles ("01" coincidence, black), without veto, 60 cm thick, blue, 5 cm thick, red, and mostly electrons ("11" coincidence, green); b) the same for the first TGE. By the yellow line, the fair-weather count rate is denoted, by 2 magenta lines – the TGE time.

In Fig.12 we show the calculated energy release histogram of the TGE electrons in the 5cm thick upper scintillator of the ASNT detector. The red curve is measured energy release in the 5 cm thick scintillator. The green curve is the histogram obtained by GEANT4 simulation using the recovered gamma ray energy spectrum and performing a full simulation of particle transport through ASNT detector. The difference between these 2 histograms, the blue line, forms the electron energy release histogram. Thus, recovered in this way number of electrons (the integral of the blue curve in Fig. 12) was ≈10400.

Most electrons release 5-10 MeV energy in the 5 cm thick scintillator, as is expected due to ≈1.8 MeV ionization losses in 1 cm of the scintillator. Abundant electrons with energies 5-10 MeV will reach the upper scintillator and leave all energy in it and didn't reach the lower scintillator. It is way the registered number of electrons in the upper 5 cm scintillator is 10400 and in the lower 60 cm thick, only 4100. These 4100 electrons were used in the energy spectrum recovery shown in Fig. 13. The higher values of energy release (> 10 MeV) are due to multiple particle traversal through the scintillator. During TGE such events are significantly enlarged compared with a fair-weather count.

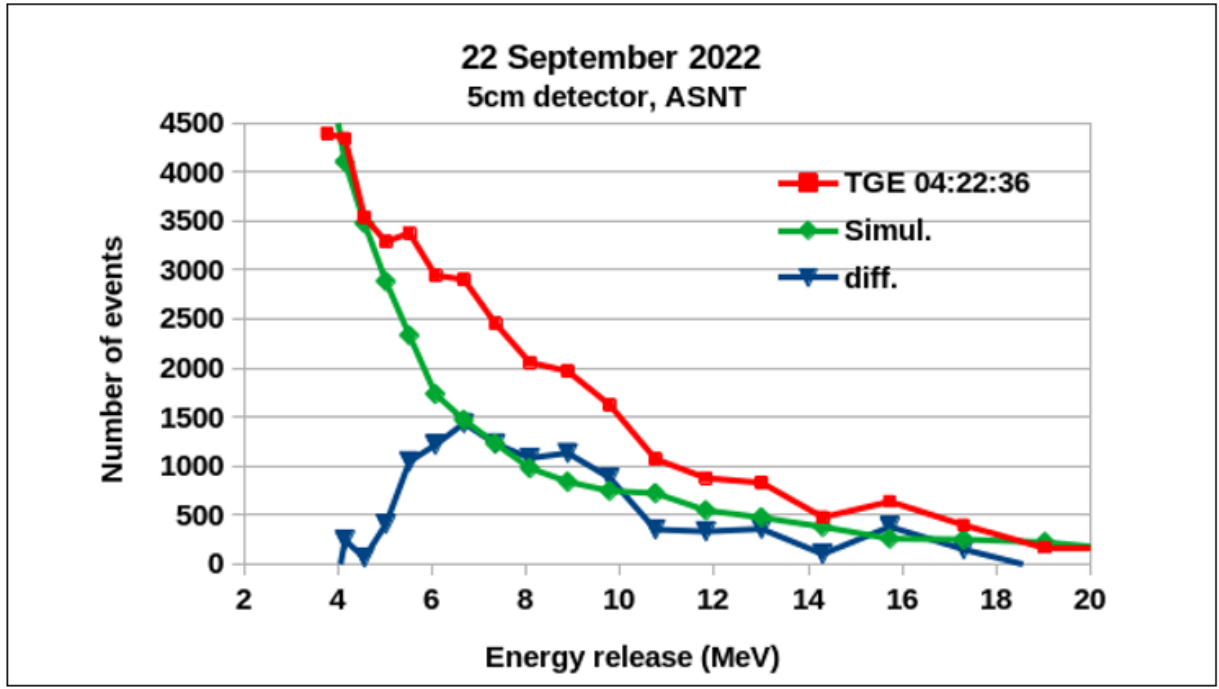

**Figure 12. Measured energy release histogram in 5cm thick scintillator at 4:22:14 – 4:23:12 UT, red; histogram of the simulated energy release of gamma rays in the same scintillator, green; difference of both, electron energy release, blue.**

The recovery of gamma ray and electron energy spectra is performed by solving the inverse problem of cosmic rays with a detailed simulation of the particle transport through the detector setup, including the selection of robust a-*priory* energy spectrum for calculating bin-to-bin migration (see details in [6]). In Fig.13 we present the resulting electron differential energy spectrum. The minimum electron energy that can be recovered by the ASNT spectrometer is 10 MeV. The fraction of electrons with energies above 10 MeV reaches ≈ 45% of the gamma ray flux. The corresponding estimate of the electric field termination height according to the approximate equation (see details in [13]) equals 70m. However, this parameter is very difficult to estimate precisely because of very fast variations in the position of charged layers in thunderclouds. We made a simulation study of the height estimation accuracy (see details in supplementary materials to [14]), and find that the standard deviation of the estimate is rather large ≈50 m.

The time of TGE in Figures 11b and 13 are slightly different because the energy release histograms are stored each 20s, and count rates each 2s. Thus, we can outline the TGE time more precisely with the 2s time series.

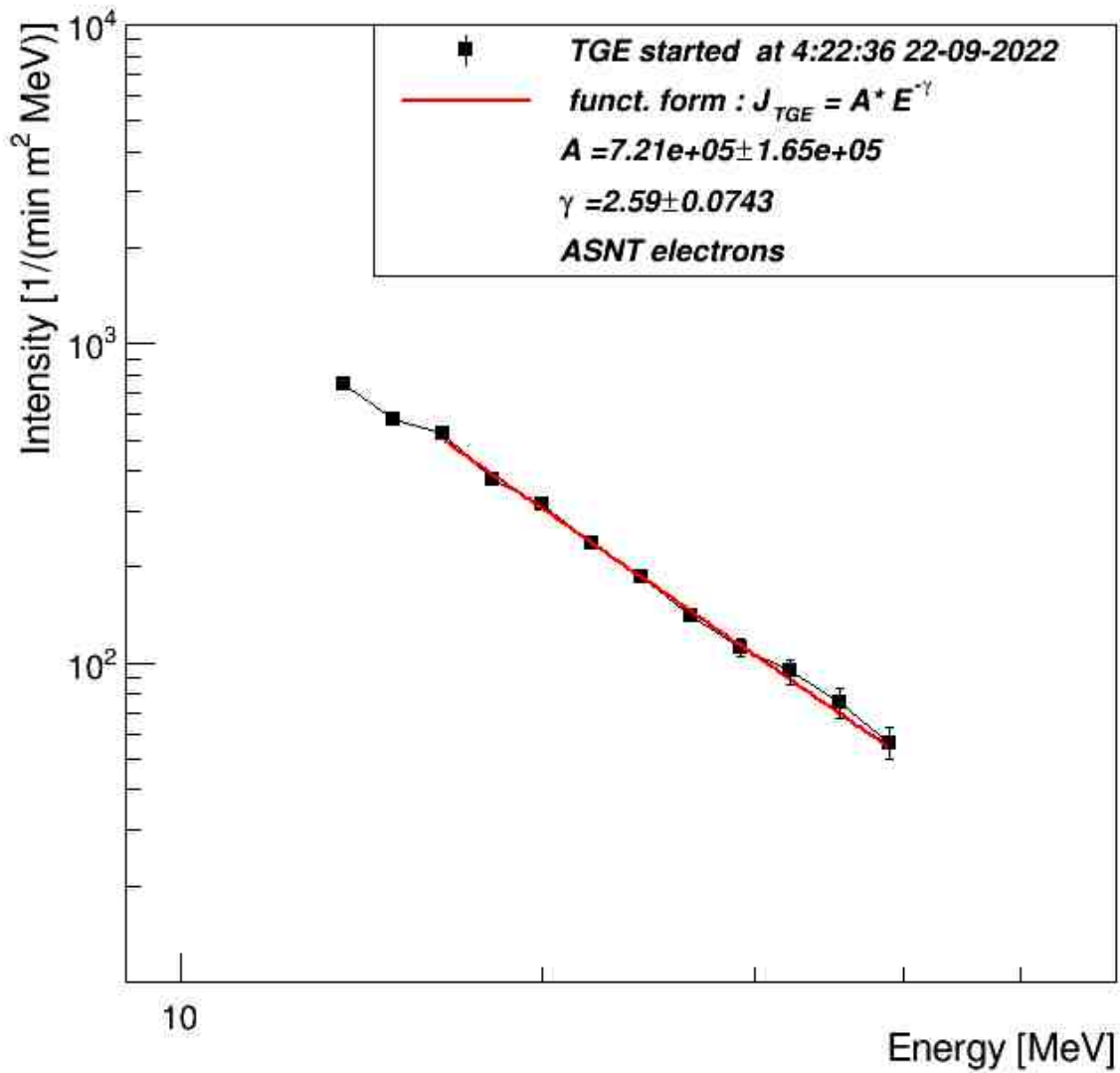

**Figure 13. Energy spectra of TGE electrons and gamma rays recovered from the energy release histograms in a 60 cm thick scintillator.**

In Fig.14 we show the comparison of the gamma ray differential energy spectra recovered with 2 independent spectrometers, ASNT and NaI. We can see the very good agreement of both spectra at energies above 4 MeV. NaI spectrometer is located under the roof built of 0.6 mm tilts and can measure energy spectrum from 0.3 MeV, however, due to the small sensitive area (0.036 m$^2$) the highest energies are measured with big errors. ASNT spectrometer measure energy spectra from 4 MeV to 100 MeV with good statistics because of a 4 m$^2$ sensitive surface. The 4 MeV energy threshold is due to a large amount of matter in the ASNT scintillators, housings, and construction elements.

With the data posted in Table 1, we can compare the spectra parameters of all 3 largest TGEs registered on 22 September. The TGE occurred at 4:22 allowed the electron energy spectrum recovery, thus, the RREA go out of the electric field region much nearer to the earth's surface than the other 2, during which the electron flux was attenuated in the dense atmosphere. The spectral knee occurred for the "electron" TGE at smaller energies than for the "far" from the ground TGEs (at ≈ 6 MeV, compared to ≈ 10 MeV); the "knee" is smoother (ε=6.9, compared to ε ~2.5), and the spectrum after the "knee" is softer: $\gamma_2$ =2.87 relative to 3.93 at maximum flux minute of THE started at 9:18.

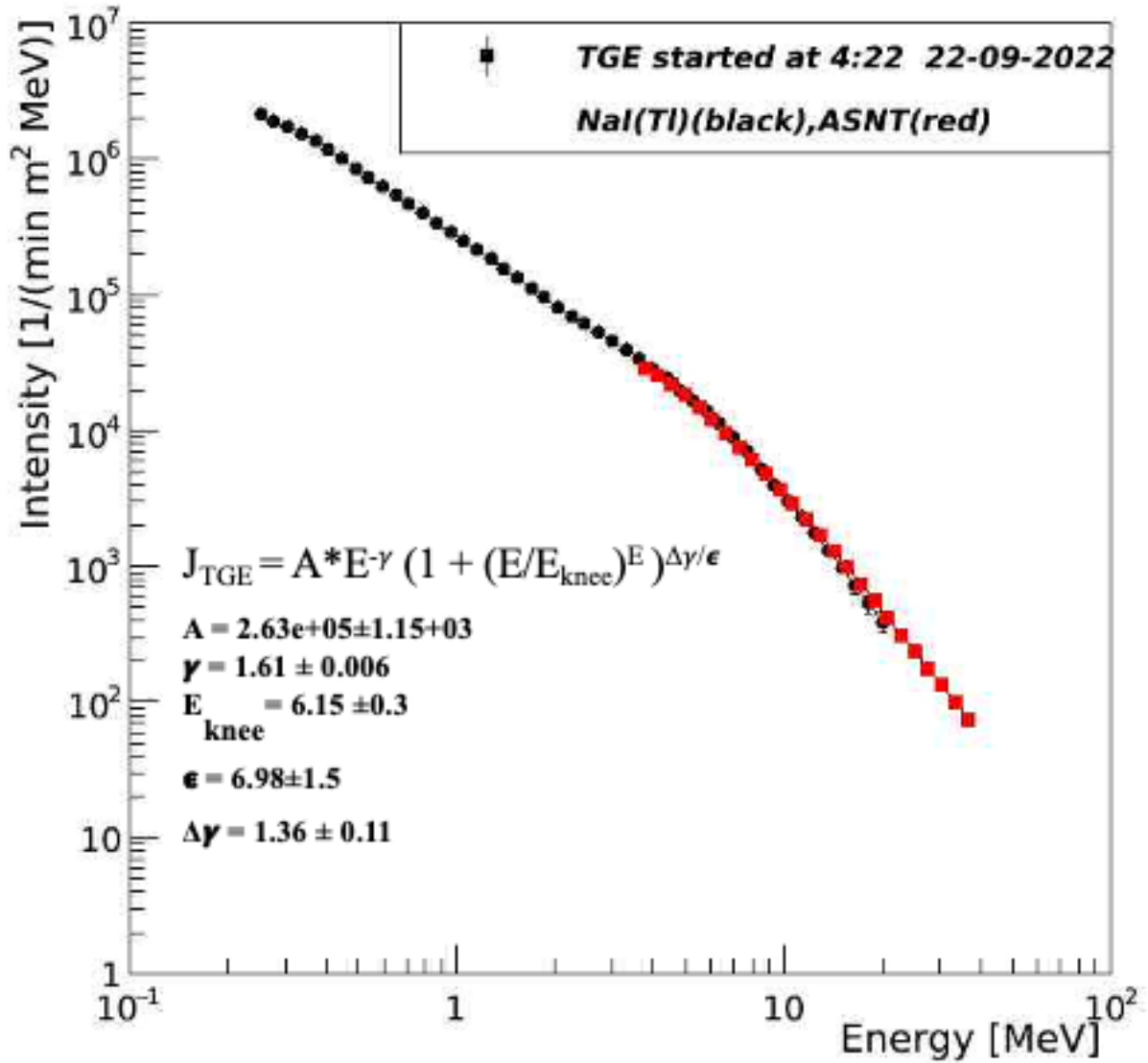

**Figure 14, Comparison of the differential energy spectra of TGE gamma rays measured with NaI (black, energy threshold 0.3 MeV, and ASNT (red, energy threshold 4 MeV) spectrometers.**

## 4. Horizontal profile of the atmospheric electric field

In Fig 15 we show a small TGE occurred at the same day later at 13:25 – 13:40. At Aragats (Fig. 15a) TGE was rather significant: ≈9 by STAND3 and ≈7 by SEVAN upper (shown in Fig. 15). TGE occurred during very intense lightning activity on Aragats, the number of nearby (within 10 km) flashes exceeds 50 in half-of-hour. In 7 minutes from 13:27 (start of TGE) to 13:35 (maximum of TGE intensity) the NSEF makes a turn from the positive +10 kV/m to the negative -15kV/m domain. As we can see in Fig.15 just after NSEF reversal we observe intense graupel fall lasting 5 minutes. Thus, LPCR sitting on graupel disappeared, and afterward, NSEF was controlled by the main negative layer in the thundercloud, which continue to maintain the lower dipole accelerating electrons downward to the earth's surface. In Fig. 15b we show the disturbances of the electric field, the 1-minute count rate of the NaI detector, and distances to lightning flashes measured at another research station, Nor Amberd, located on slopes of Mt. Aragats 1200 m lower than Aragats station, and ≈13 km apart. The TGE event observed in Nor Amberd was also rather significant exceeding 11 . Thus, although the pattern of the NSEF disturbances was rather different at both locations, the intracloud electric field was strong enough to unleash the runaway process and send numerous electrons and gamma rays in direction of the earth's surface. It was another proof (see for the first detection of the extended TGE flux in [7]), that during thunderstorms large areas below thunderclouds are radiated by the electron accelerator operated in the lower dipole.

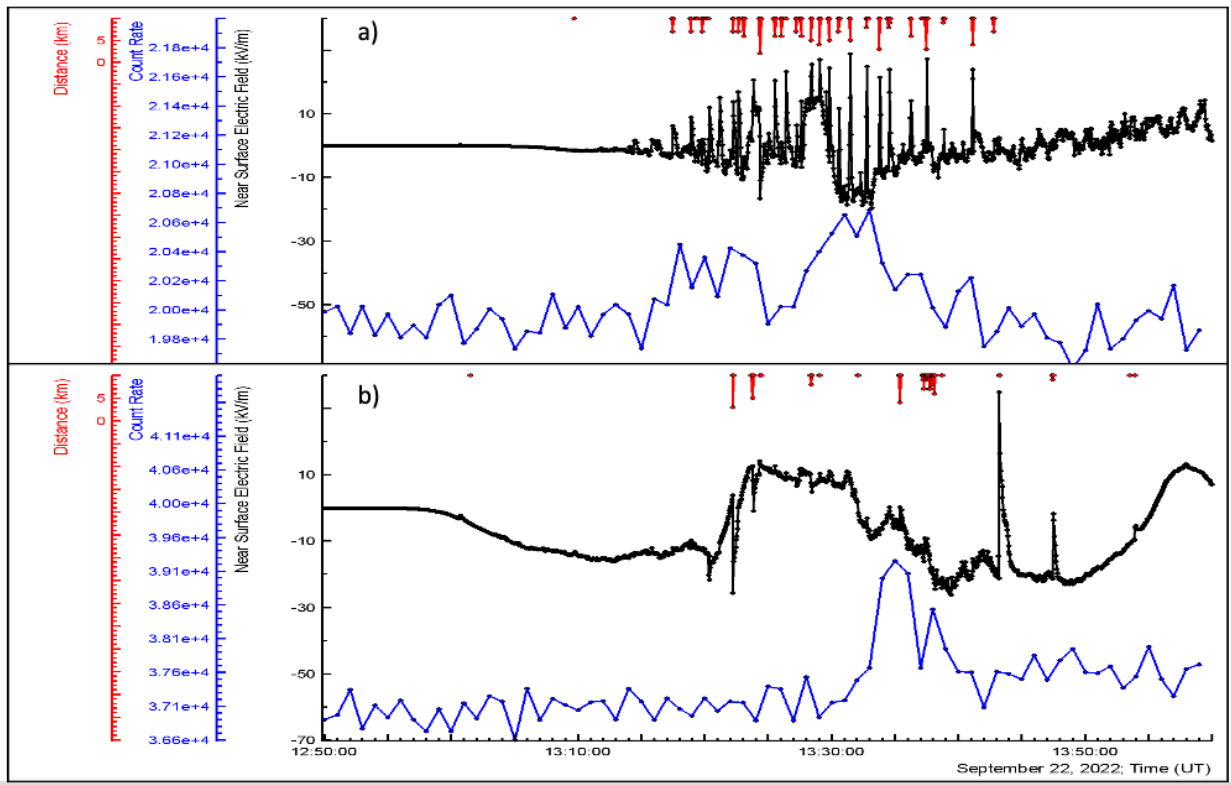

**Figure 15. a) Aragats (3200 a.s.l.), disturbances of the NSEF, black; 1minute time series of SEVAN detector's count rate, blue, and distances to lightning flashes, red; b) Nor Amberd (2000 a.s.l), the same, count rate of NaI detector.**

In Fig. 16 we show shots of the panoramic camera, which is monitoring the skies above Aragats. We can see characteristic specks on the glass of the camera, which were identified by station staff as conical graupel [10].

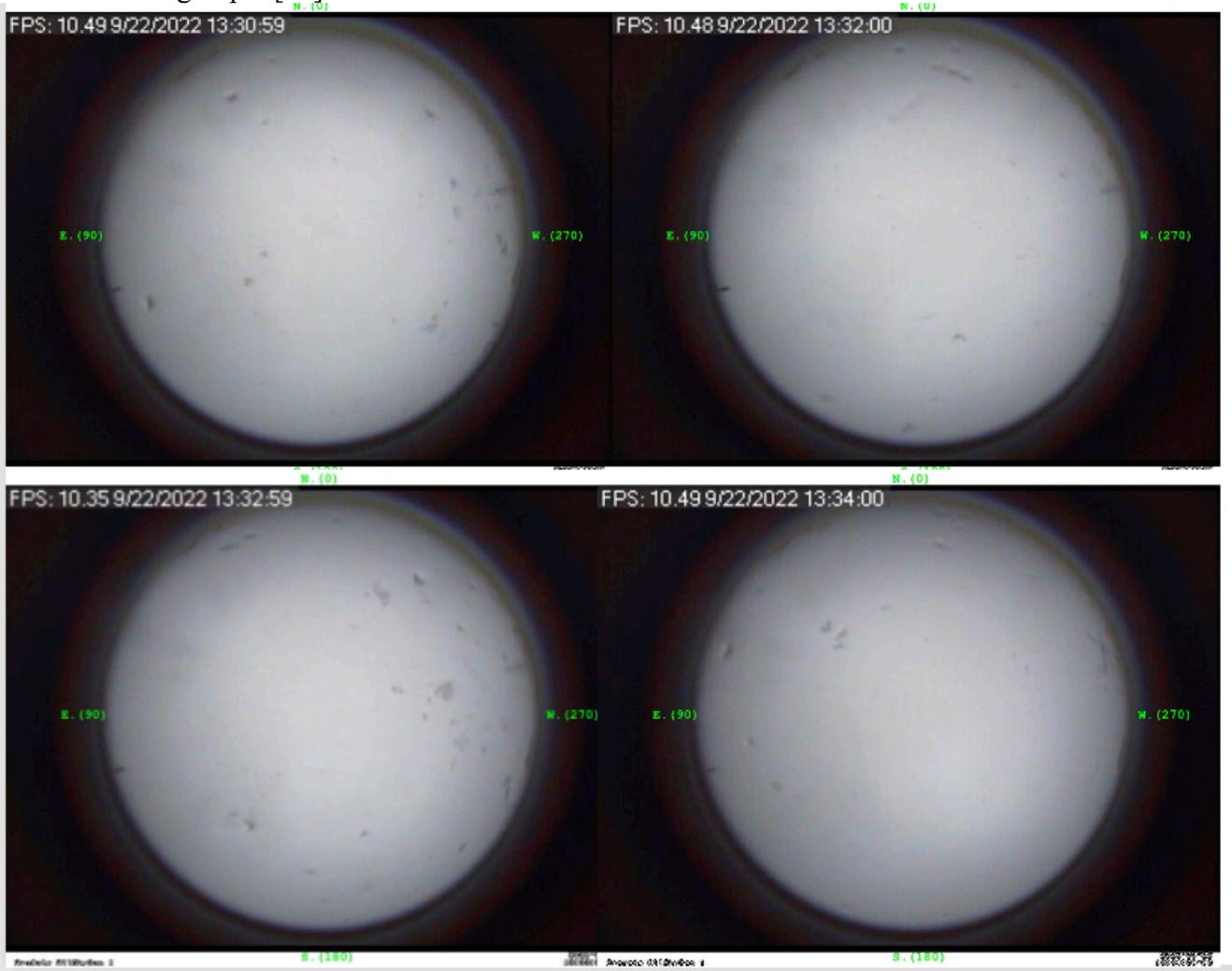

Figure 16. Shots of the panoramic camera with characteristic specks on the glass, identified as graupel fall.

In Nor Amberd the NaI detector was surrounded by two $0.8m^2$ and 5 cm thick plastic scintillators, occupying a flat area of 2 square meters. The coincidence of these 3 detectors highly enhanced during TGE, from the fair-weather mean of ≈20 to 64, see Fig. 17. The enhancement started at 13:34 with a sign reversal of the NSEF. The large enhancement continued for 5 minutes to 13:39, when NSEF was in the deep negative domain with a minimum of -25 kV/m. Thus, the RREA avalanches hitting 3 detectors, started when LPCR was destroyed and NSEF undergo the sign reversal.

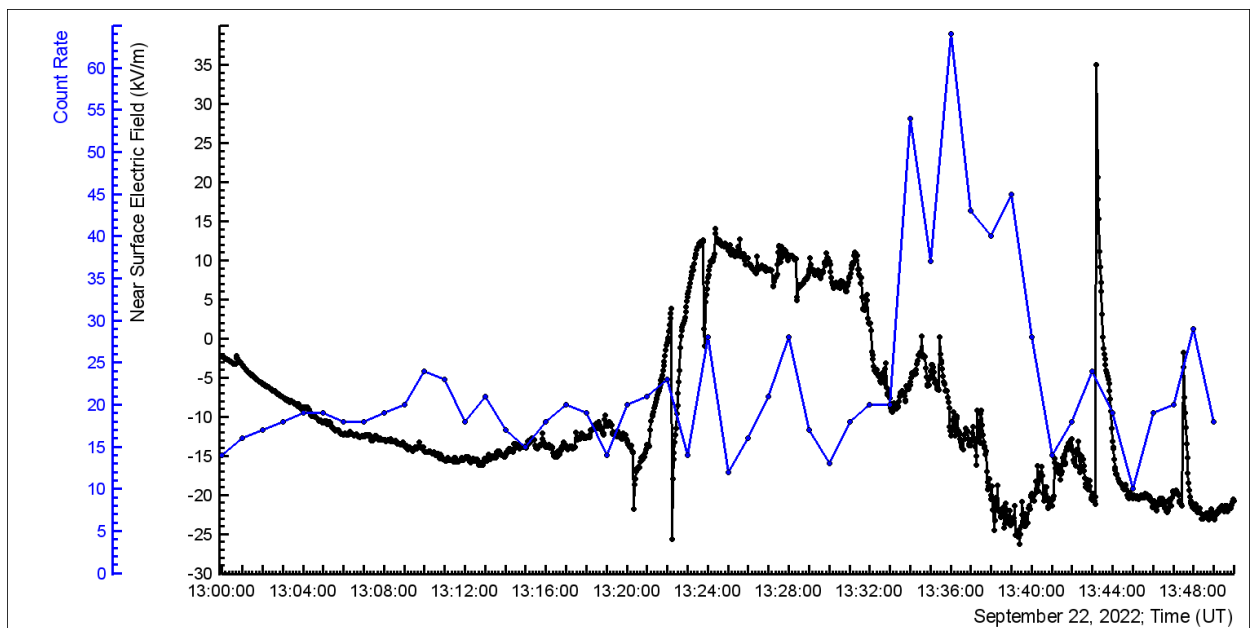

**Figure 17.** Disturbances of NSEF, black, and coincidence count rate from 3 detectors: 2 plastic scintillators and NAI crystal, blue.

5. **Small TGE events occurred on 22 September 2022: the LPCR emergence and contraction.**

The TGE event that occurred at 3:10-3:22 was not large by amplitude but extended 12 minutes, allowing to follow excursions of NSEF and corresponding changes of the particle count rates and lightning flashes. The first TGE started during positive NSEF of +8kV/m was abruptly terminated after ≈1 minute by an inverted polarity flash stroked at a ≈1.5 km distance.

Attempt to start TGE afterward during negative NSEF reaching -8 kV/m at 3:12-3:16 were unsuccessful, and count rate enhancement was very small. At 3:17-3:19, again sign reversal of NSEF occurred recovering +8 kV/m value. However, 2 nearby normal polarity flashes (at distances 6.5 and 4.5 km) did not allow TGE development. Only after another NSEF sign reversal when the NSEF value reaches -18 kV/m does TGE start to develop at 3:19 and smoothly finished at 3:22, see Fig. 18. Thus, we observe TGEs both at negative and positive NSEF.

Attempts to start TGE during not large negative NSEF (-9 kV/m) were not successful, only during deep negative NSEF count rate start to significantly enhance.

During positive NSEF, an LPCR was formed, as we can see from the patterns on the glass of the panoramic camera in Fig. 19. After finishing both episodes of positive NSEF we detect the graupel fall (3:12-3:13, and 3:18-3:20), which evidenced the decay of the LPCR sitting on graupel.

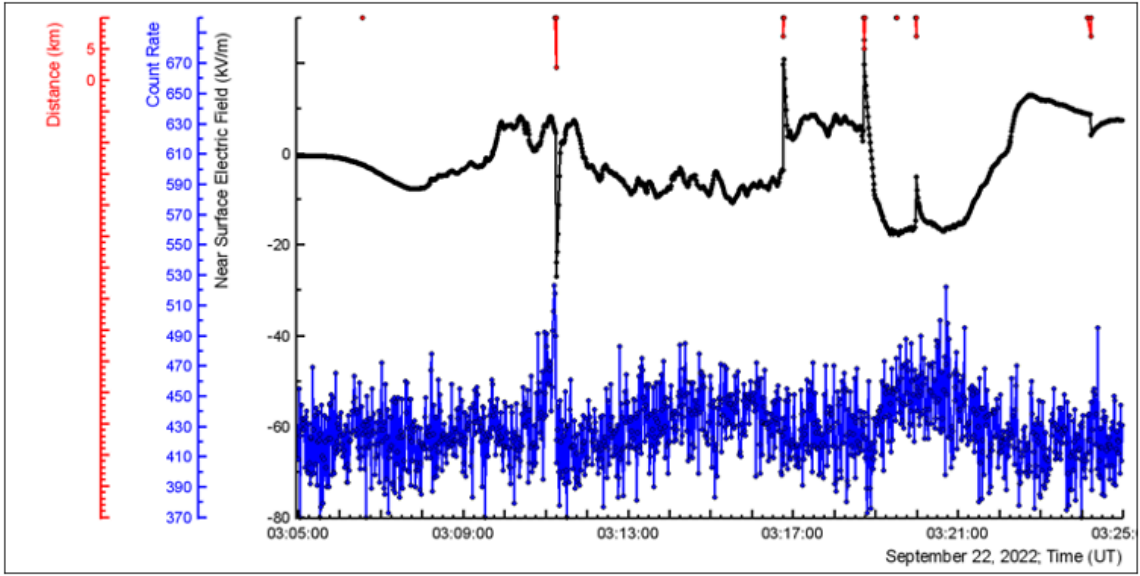

**Figure 18.** Disturbances of the NSEF, black; time series of 1s count rates of STAND1 plastic scintillator of 1 m$^2$ area and 1 cm thickness, blue; distances to lightning flashes, red.

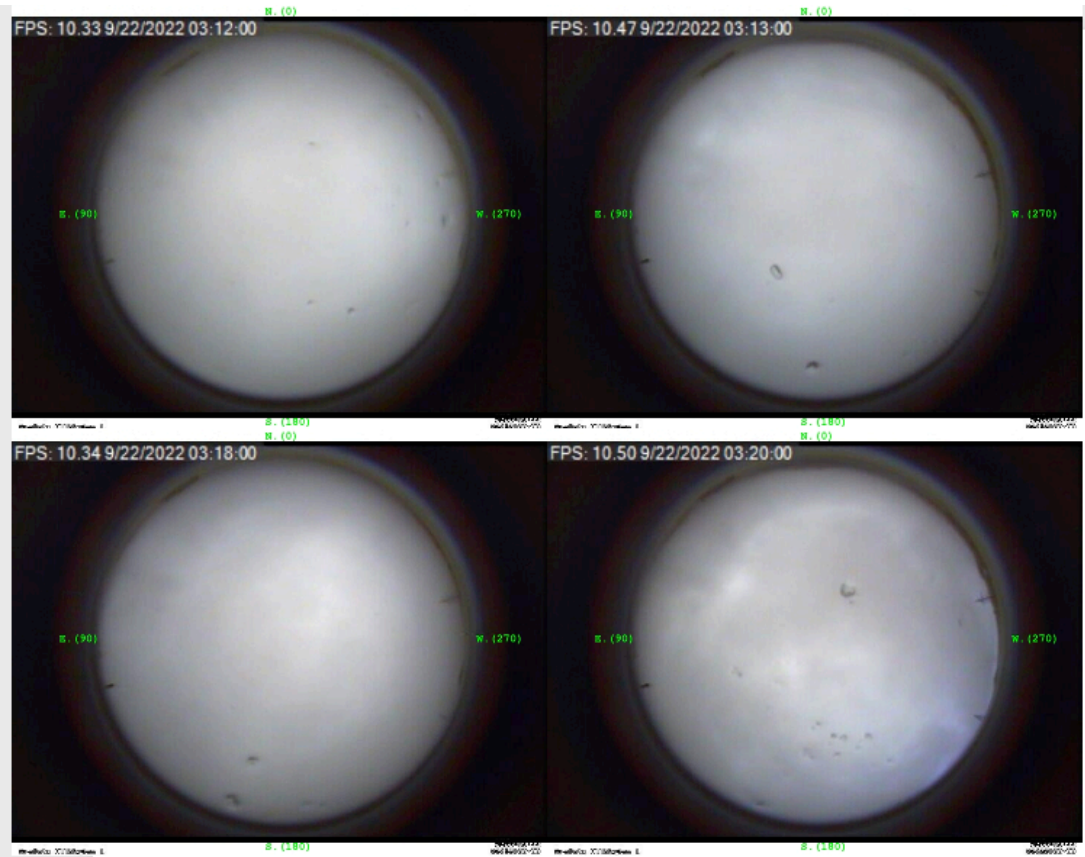

**Figure 19.** Shots of the panoramic camera showing characteristic specks on the glass are identified with graupel fall.

Another small TGE occurred again during the positive NSEF that reaches +18 kV/m and extended from 4:57 to 5:07. The first attempt to start TGE at 4:59 was terminated by inverted polarity lightning. However, afterward, TGE restarted at 5:00 and smoothly finished at 5:05, see Fig. 20. The camera shots again demonstrate characteristic specks of the graupel fall at 5:07-5:08, just after decaying of the positive NSEF and contracting of TGE, see Fig. 21. Thus, for this TGE, lower dipole was formed by the main negative and LPCR charged layers and decay of the LPCR with fallen graupel terminates TGE.

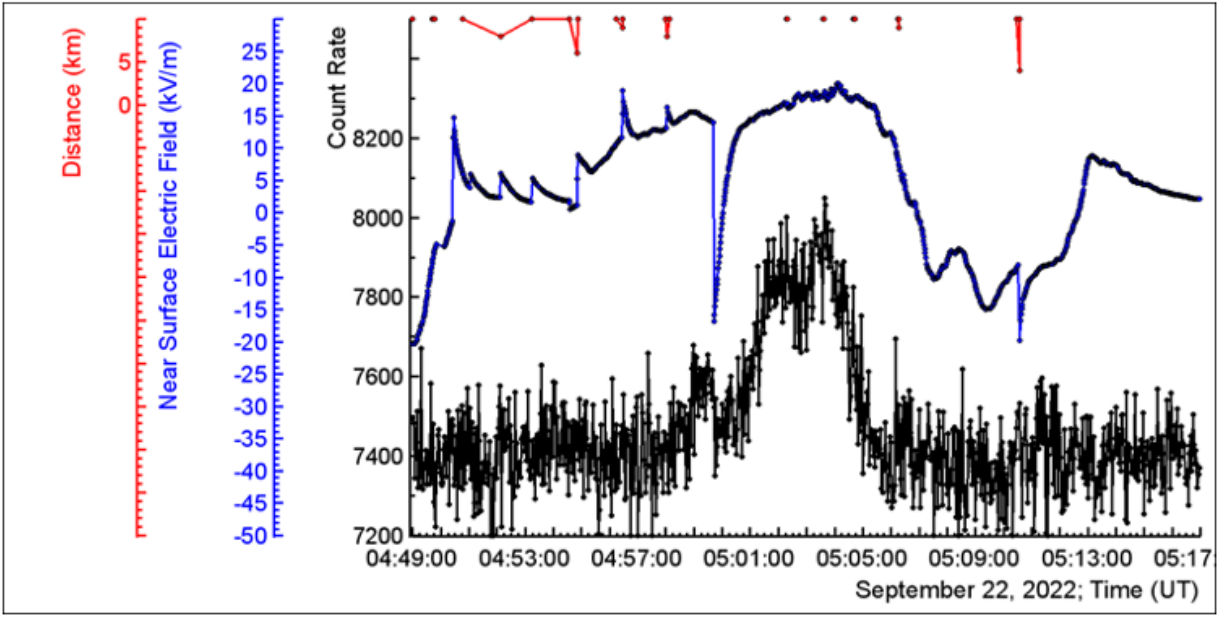

**Figure 20. Time series of 2s count rates of ASNT scintillator of 4 m² area and 60 cm thickness, black; disturbances of the NSEF, blue; distances to lightning flashes, red.**

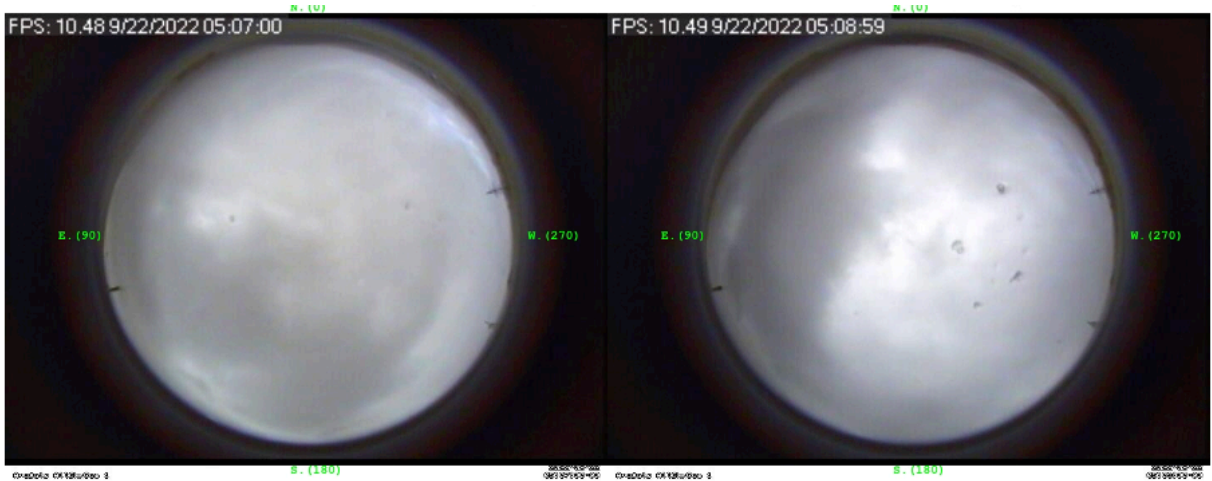

**Figure 21. Shots of the panoramic camera showing characteristic specks on the glass are identified with graupel fall.**

## 6. Conclusion

On September 22, 2022, nature demonstrates a variety of TGE events each of a specific flavor. Using the developed in last decade methodology of TGE analysis [15], we show that TGEs occurred during positive NSEF, during negative NSEF, and also, when the electric field undergoes sign reversal. Correspondingly, la ower dipole was formed by the main negative charge region in the middle of the cloud and its mirror in the earth, or by the same main negative and LPCR. We demonstrate by multiple graupel fall patterns that LPCR is a transient layer sitting on graupel and disappeared with graupel fall.

We show that RREA can cover many cubic km in the thundercloud, and then when finish as TGE – many square kilometers on the earth's surface.

On 22 September, unlike TGE events described in our previous paper when the lightning active zone miss the Aragats station [16], the active lightning zone was just above the stations, and normal and inverted polarity lightning flashes abruptly terminated TGEs. Thus, it is evidence that RREAs/TGEs are precursors of lightning flashes [17].

Only one TGE from 5 discussed in the papers allows recovery of the electron energy spectrum, we estimate the height of a strong accelerated electric field above spectrometers to be ≈70m. Gamma ray energy spectrum measured at minute 9:19-9:20 was extremely intensive and energetic. TGE during the most intense minute (from four) sends 1.25 mln particles (with energies exceeding 0.3 MeV) per minute per $m^2$, the highest energy reaching 70 MeV. However, at highest energies the contribution of the MOS process [10], negligible at lower energies, becomes essential. Thus, the TGE particle energies can be recovered reliably up to 50 MeV.

## Acknowledgements

We thank the staff of the Aragats Space Environmental Center for the uninterruptable operation of all particle detectors and field meters. The authors acknowledge the support of the Science Committee of the Republic of Armenia (research project № 21AG-1C012), in the modernization of the technical infrastructure of high-altitude stations.

**Data availability statement**: The data that support the findings of this study are openly available at the following URL[database]: http://adei.crd.yerphi.am/.